\documentclass[aps,prl,reprint,groupedaddress]{revtex4-2}

\usepackage[dvipdfmx]{graphicx}
\usepackage{amssymb}
\usepackage{dcolumn}
\usepackage{bm}
\usepackage{color}
\usepackage{xcolor}
\usepackage{amsmath}
\usepackage{amsfonts}
\usepackage{braket}
\usepackage{cancel}

\DeclareMathOperator{\sech}{sech}
\begin{document}

\title{Analog black-white hole solitons in traveling wave parametric amplifiers with superconducting nonlinear asymmetric inductive elements}

\author{Haruna Katayama$^{1,2}$}  
\email[]{halna496@gmail.com.}
\author{Noriyuki Hatakenaka$^{2}$}
\author{Toshiyuki Fujii$^{3}$}
\author{Miles P. Blencowe$^{1}$}

\affiliation{$^{1}$Department of Physics and Astronomy, Dartmouth College, Hanover, New
Hampshire 03755, USA\\
$^{2}$Graduate School of Advanced Science and Engineering, Hiroshima University, Higashihiroshima, Hiroshima 739-8521, Japan\\
$^{3}$Department of Physics, Asahikawa Medical University, Midorigaoka-higashi, Asahikawa, Hokkaido 078-8510, Japan}

\begin{abstract}
We show that existing travelling wave parametric amplifier (TWPA) setups, using superconducting nonlinear asymmetric inductive elements (SNAILs), admit soliton solutions that act as analogue event horizons. 
The SNAIL-TWPA circuit dynamics are described by the Korteweg-de Vries (KdV) or modified Korteweg-de Vries (mKdV) equations in the continuum field approximation, depending on the external magnetic flux bias, and validated numerically. The soliton spatially modulates the velocity for weak probes, resulting in the effective realization of analogue black hole and white hole event horizon pairs. The SNAIL external magnetic flux bias tunability facilitates a three-wave mixing process, which enhances the prospects for observing Hawking photon radiation.
\end{abstract}

\maketitle

\newpage

\emph{Introduction.}---
Analogue black holes have been proposed in various laboratory systems as a means of verifying the basic principle of Hawking radiation \cite{Hawking1975}, which could provide a clue for unifying gravity and quantum mechanics. 
In 1981, Unruh pioneered the study of analogue black holes by showing the analogy between the behavior of sound waves in a transsonic fluid flow and that of light in the spacetime of a black hole \cite{Unruh1981}. The basic idea is to create a spatially varying fluid flow so that sound waves cannot escape from a certain boundary corresponding to a black hole event horizon (i.e., a sonic event horizon). Based on this idea, thermal properties of stimulated Hawking emission and the correlations that are associated with the Hawking effect have been observed in a water flume \cite{Weinfurtner2011,Euve2016}. But such experiments cannot capture the quantum spontaneous emission aspects of Hawking radiation due to the overwhelming classical thermal noise.

In order to observe quantum effects, analogue black holes have been proposed in various systems such as Bose-Einstein condensates \cite{Steinhauer2014,Steinhauer2016}, optical fibers \cite{Philbin2008,Choudhary2012,Drori2019}, and superfluids in polariton microcavities \cite{Nguyen2015,Jacquet2020}. Cryogenic, superconducting transmission line circuits have advantages over the above schemes in controllability, scalability, and low noise operation, making the detection of quantum correlated Hawking radiation involving  {\it photons} a very real possibility \cite{Nation2012,Wilson2011,Lahteenmaki2012}. A spatially varying microwave velocity is a necessary requirement for creating an analogue black hole in such a circuit \cite{Schutzhold2005,Nation2009,Katayama2020, Katayama2021,Katayama2021ieee,Katayama2021universe,Kogan2022}. This can be achieved by introducing an effective spatially dependent inductance $L$ or capacitance $C$, since the electromagnetic wave velocity is given by $v=a/\sqrt{LC}$ with unit cell length $a$. The problem of heating that hinders the observation of Hawking radiation \cite{Schutzhold2005} has been addressed by using superconducting circuits \cite{Nation2009}. Additionally, the issue of pulse instability arising from dispersion has been overcome by utilizing solitons \cite{Katayama2020}, leading to classically stable horizons.
However, there is still some way to go in terms of realizing the design, fabrication and measurements in order to successfully observe Hawking radiation. 
One unexpected obstacle is the Kerr effect itself inherent in conventional Josephson systems, i.e., the fourth-order nonlinearity of the Josephson effect required for soliton formation as well as the Hawking pair creation from the vacuum.
The Kerr interaction causes higher-order harmonic generation due to four-wave mixing, which reduces the degree of entanglement and squeezing performance essential for Hawking radiation detection \cite{Boutin2017,Peng2022}.

\begin{figure}[htbp]
\centering
\includegraphics[width=8cm]{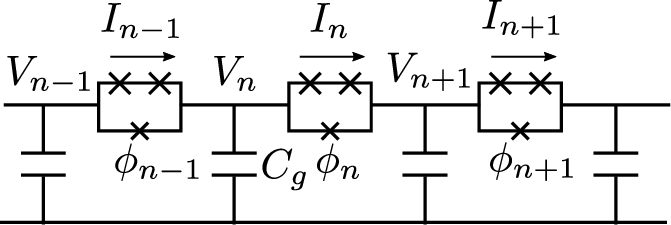}
\caption{Schematic diagram of the traveling wave parametric amplifier (TWPA) transmission line with superconducting nonlinear asymmetric inductive element (SNAIL) unit cells in series, alternating with shunt capacitors $C_g$ in parallel (all assumed identical). The quantities $I_n$, $V_n$, and $\phi_n$ represent the current, voltage, and phase difference of the $n$th SNAIL, respectively. Each SNAIL has a small JJ with Josephson energy $\alpha E_J$ ($\alpha<1$) in one branch and two larger JJs (with Josephson energy $E_J$) in the other, parallel branch, forming a loop that is threaded by an external applied magnetic flux $\Phi_{{\mathrm{ext}}}$. %(a) `(S)uperconducting (N)onlinear (A)symmetric (I)nductive e(L)ement' (`SNAIL'), which has a small JJ with Josephson energy $\alpha E_J$ ($\alpha<1$) in one branch and two larger JJs (with Josephson energy $E_J$) in the other, parallel branch, forming a loop that is threaded by an external applied magnetic flux $\Phi_{{\mathrm{ext}}}$. The SNAIL is indicated schematically by a spiral, reflecting the physical geometry of the fabricated device \cite{Frattini2017}, and the phase difference across the SNAIL element is represented by the $\phi$ coordinate. %(b) Schematic diagram of the `(T)raveling (W)ave (P)arametric (A)mplifier' (`TWPA') transmission line with SNAIL unit cells in series, alternating with shunt capacitors $C_g$ in parallel (all assumed identical). The quantities $I_n$, $V_n$, and $\phi_n$ represent the current, voltage, and phase difference of the $n$th SNAIL, respectively.
} \label{fig:model}
\end{figure}

A three-wave mixing process caused by a third-order nonlinear potential can improve these aspects. 
However, the ordinary Josephson effect {\it alone} cannot produce odd-order nonlinearities such as a cubic potential.
One way to achieve this is through a superconducting nonlinear asymmetric inductive element (SNAIL) \cite{Frattini2017}, which is a superconducting loop consisting of a single small Josephson junction (JJ) and several larger junctions in parallel as shown in Fig. \ref{fig:model}. This results in a third-order nonlinear effect, in addition to the leading fourth-order nonlinear effect of a single JJ. Travelling wave parametric amplifier (TWPA) transmission line devices incorporating SNAILs have demonstrated remarkable results in recent experiments \cite{Esposito2022,Perelshtein2022}. Specifically, parametrically-generated multimode entangled microwave photons were successfully observed, which highlights the promising prospect for generating and detecting Hawking radiation in such devices. However, it is not a priori obvious whether such a third-order nonlinearity can %contribute to creating
give rise to a soliton with an analogue black hole event horizon.
%However, it is \blue{nontrivial} whether the third\blue{-order} nonlinearity can \blue{contribute} to create an analogue black hole \blue{soliton}.

%In this system, the four-wave mixing process is caused by the Kerr nonlinearity, where the soliton behaves as a pump, and Hawking radiation and its partner play the role of signal and idler, respectively. However, it is not favorable for observing the Hawking radiation because the Hawking particle and its partner have close frequencies to that of the pump.

% in this paper
In this letter, we propose an analogue black hole soliton realization based on a Kerr-free nonlinearity using a transmission line comprising %superconducting nonlinear asymmetric inductive element (SNAIL) 
SNAIL unit cells. %\cite{Frattini2017}.
In particular, we shall establish both analytically and numerically the existence of soliton solutions to the nonlinear transmission line dynamical equations for the SNAIL phase difference coordinates. %in the spatial continuum approximation, by using the reductive perturbation method \cite{Taniuti1968,Taniuti1969,Taniuti1974}. The third and fourth-order nonlinearities lead respectively to the Korteweg-de Vries (KdV) \cite{KdV1895} and modified Korteweg-de Vries (mKdV) \cite{Miura1968,Wadati1973} equations, which are well-known to have exact, solitary wave (i.e., `soliton' in short)  solutions \cite{kivshar1989}.  A stably propagating soliton is formed in the superconducting circuit transmission line by balancing the dispersion and nonlinearity of the SNAIL unit cell elements \cite{Katayama2020,Katayama2021ieee,Katayama2021,Katayama2021laser} (see Refs. \cite{ustinov1998,wildermuth2022} for related work on solitons in JJ devices). 
%We also numerically confirm that the analytically obtained soliton solutions propagate in the original, discrete circuit equations of the SNAIL TWPA. 
These obtained solitons spatially modulate the velocity of weak `probes' %microwaves 
and two stable horizons occur  where the probe %microwave 
and soliton velocities coincide, resulting in an analogue black hole and white hole horizon pair. Neglecting (quantum) noise and dissipation in the circuit dynamics, a soliton will propagate along the transmission line without broadening, in contrast to a background `pulse' solution to the linear phase coordinate wave equations, which undergoes dispersion \cite{Nation2009}. This stability has the advantage that the effective Hawking temperature (which depends on the probe velocity gradient at the effective horizon) does not decrease as the soliton propagates (with backreaction neglected).   

%The remainder of this paper is organized as follows. We begin with a description of the SNAIL TWPA model and derive its circuit equations. We then discuss both the strong background wave pulse and weak probe signal solutions in the presence of the strong background pulse.  By using the reductive perturbation method and numerical simulations, we show that the strong background wave pulse can be a soliton solution. %for appropriate initial conditions. 
%In the final part, we show that the background soliton modulates the velocity of the probe signal and creates an analogue black-white hole event horizon pair. The paper ends with a summary. 

%The work contained in the present paper sets the stage for analyzing the quantum microwave Hawking radiation in the background effective spacetime of the soliton pulse solutions; this will be the subject of a follow-up investigation \cite{followup}. 

\begin{figure}[htbp]
\centering
\includegraphics[width=7cm]{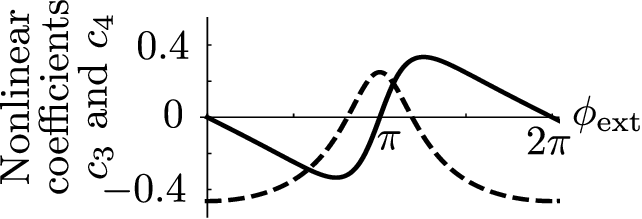}
\caption{The  external flux $\phi_{{\mathrm{ext}}}$ dependence of the respective third and fourth order nonlinear coefficients $c_3=\tilde{\beta}\left(\phi_{{\mathrm{ext}}}\right)/\tilde{\alpha}\left(\phi_{{\mathrm{ext}}}\right)$ (solid line) and $c_4=\tilde{\gamma}\left(\phi_{{\mathrm{ext}}}\right)/\tilde{\alpha}\left(\phi_{{\mathrm{ext}}}\right)$ (dashed line) of a SNAIL.} \label{fig:nonlinearity}
\end{figure}

%======================
%\section{Model}
\emph{Model.}---Consider a transmission line with SNAILs %which are superconducting loops comprising a `small' JJ and two `large' JJ's 
as shown in Fig. \ref{fig:model}, where the Josephson energy of the small and large junctions are $\alpha E_J$ and $E_J$, respectively (with $\alpha<1$), and all capacitors have the identical capacitance $C_g$ in the shunt branch. In the following, we consider the circuit equations of the transmission line, beginning first with the potential energy of a single SNAIL, which is given as 
\begin{align}
U\left(\phi\right)&=-\alpha E_J\cos\left(\phi\right)-2E_J\cos\left(\frac{\phi-\phi_{{\mathrm{ext}}}}{2}\right)+{\mathrm{const.}},\nonumber\\
&\simeq\!E_J\left[\!\frac{\tilde{\alpha}(\phi_{{\mathrm{ext}}})}{2!}{\tilde{\phi}}^2+\frac{\tilde{\beta}(\phi_{{\mathrm{ext}}})}{3!}{\tilde{\phi}}^3+\frac{\tilde{\gamma}(\phi_{{\mathrm{ext}}})}{4!}{{\tilde{\phi}}}^4\!\right].
\end{align}
Here, $\phi$ is the phase difference across the smaller junction and $\phi_{{\mathrm{ext}}}=2 \pi\Phi_{{\mathrm{ext}}}/\Phi_{0}$, with $\Phi_{{\mathrm{ext}}}$  and $\Phi_{0}=h/(2e)$  the (tunable) external magnetic flux bias and magnetic flux quantum, respectively, and the Taylor expansion around  the local minimum $\phi^\ast$ of the potential is performed, %as follows:
%\begin{align}
%U\left(\phi^\ast+\tilde{\phi}\right)=& U\left(\phi^\ast\right)+\left.\ \frac{dU}{d\phi}\right|_{\phi=\phi^\ast}\tilde{\phi}+\left.\ \frac{1}{2!}\frac{d^2U}{d\phi^2}\right|_{\phi=\phi^\ast}{\tilde{\phi}}^2\nonumber\\
%&+\left.\ \frac{1}{3!}\frac{d^3U}{d\phi^3}\right|_{\phi=\phi^\ast}{\tilde{\phi}}^3+\left.\ \frac{1}{4!}\frac{d^4U}{d\phi^4}\right|_{\phi=\phi^\ast}{\tilde{\phi}}^4\nonumber\\
%&+\mathcal{O}\left({\tilde{\phi}}^5\right),\end{align}
with $\tilde{\phi}$ being the variation about $\phi^\ast$ and setting $U\left(\phi^\ast\right)=dU/d\phi|_{\phi=\phi^\ast}=0$. %, we obtain
%\begin{align}
%U(\phi)\!=\!E_J\left[\!\frac{\tilde{\alpha}(\phi_{{\mathrm{ext}}})}{2!}{\tilde{\phi}}^2+\frac{\tilde{\beta}(\phi_{{\mathrm{ext}}})}{3!}{\tilde{\phi}}^3+\frac{\tilde{\gamma}(\phi_{{\mathrm{ext}}})}{4!}{{\tilde{\phi}}}^4\!\right],
%\end{align}
%where the external magnetic flux dependent coefficients are defined as follows:
%\begin{align}
%\tilde{\alpha}(\phi_{{\mathrm{ext}}})&=\left.\ \frac{1}{E_J}\frac{d^2U}{d\phi^2}\right|_{\phi=\phi^\ast}\nonumber\\
%&=\alpha \cos \phi^{*}+\frac{1}{2} \cos \left(\frac{\phi^{*}-\phi_{{\mathrm{{\mathrm{ext}}}} }}{2}\right),\\
%\tilde{\beta}(\phi_{{\mathrm{{\mathrm{ext}}}}})&=\left.\ \frac{1}{E_J}\frac{d^3U}{d\phi^3}\right|_{\phi=\phi^\ast}\nonumber\\
%&=-\alpha \sin \phi^{*}-\frac{1}{4} \sin \left(\frac{\phi^{*}-\phi_{{\mathrm{ext}} }}{2}\right),\\
%\tilde{\gamma}(\phi_{{\mathrm{ext}}})&=\left.\ \frac{1}{E_J}\frac{d^4U}{d\phi^4}\right|_{\phi=\phi^\ast}\nonumber\\
%&=-\alpha \cos \phi^{*}-\frac{1}{8} \cos \left(\frac{\phi^{*}-\phi_{{\mathrm{ext}}}}{2}\right).
%\end{align}
In contrast to a single JJ, the SNAIL has the advantage of a non-zero, odd-order nonlinearity because of quantum interference. Figure \ref{fig:nonlinearity} gives the $\phi_{{\mathrm{ext}}}$ flux dependence of these coefficients normalized by $\tilde{\alpha}\left(\phi_{{\mathrm{ext}}}\right)$ [i.e.,   $c_3=\tilde{\beta}\left(\phi_{{\mathrm{ext}}}\right)/\tilde{\alpha}\left(\phi_{{\mathrm{ext}}}\right)$, $c_4=\tilde{\gamma}\left(\phi_{{\mathrm{ext}}}\right)/\tilde{\alpha}\left(\phi_{{\mathrm{ext}}}\right)$]. The coefficients of the third and fourth-order nonlinearities are anticorrelated, and either can be selectively set to zero by tuning $\phi_{\mathrm{ext}}$. Thus, we can choose either a three or four-wave mixing process induced by the third or fourth-order nonlinearities, respectively, 
%an advantage of SNAILs is its ability to switch between a three and four-wave mixing process induced by the third and fourth-order nonlinearities, respectively, 
by controlling the external magnetic flux in situ, without the need to alter the circuit design hardware. 

%circuit equation

With the approximate potential energy of the SNAIL in hand, we can write down the circuit equations for the TWPA. %Starting with 
From Kirchhoff's current conservation law and the Josephson voltage relation, %, we have 
%\begin{align}
%I_{n+1}-I_n=C_g\frac{dV_{n+1}}{dt},    
%\end{align}
%and combining with the voltage Josephson relation 
%\begin{align}
%V_{n+1}-V_n=\frac{\hbar}{2e}\frac{d\phi_n}{dt},\label{eq:Josephson-relation}
%\end{align}
%we obtain 
%\begin{align}
%I_{n+1}-2I_n+I_{n-1}=\frac{C_g\hbar}{2e}\frac{d^2\phi_n}{dt^2},\label{eq:Kirchhoff}
%\end{align}
%where $I_n$ %, $V_n$, 
%and $\phi_n$ are the current %, voltage 
%and phase difference of the $n$th SNAIL, respectively. 
%The current through the $n$th SNAIL is given as the sum of the displacement current and Josephson current:
%\begin{align}
%I_n=&\frac{C_J\hbar}{2e}\frac{d^2\phi_n}{dt^2}\nonumber\\
%&+I_c\left[\tilde{\alpha}\left(\phi_{\mathrm{ext}}\right)\phi_n+\frac{\tilde{\beta}\left(\phi_{\mathrm{ext}}\right)}{2!}{\phi_n}^2+\frac{\tilde{\gamma}\left(\phi_{\mathrm{ext}}\right)}{3!}{\phi_n}^3\right],\label{eq:nth_current}
%\end{align}
%with $C_J$ the Josephson capacitance  and $I_c$ the critical current  of the larger JJ in the SNAIL. Substituting Eq. \eqref{eq:nth_current} into Eq. \eqref{eq:Kirchhoff}, 
we obtain  the following circuit equations:
\begin{align}
&\frac{d^{2}\phi_{n}}{d t^{2}}-r\frac{d^{2}}{d t^{2}}\left(\phi_{n+1}-2 \phi_{n}+\phi_{n-1}\right)\nonumber \\
&-\omega_{0}^{2}\sum^{3}_{j=1}\left[\frac{c_{j+1}}{j!}\left(\phi^j_{n+1}-2 \phi^j_{n}+\phi^j_{n-1}\right)\right]=0,\label{eq:circuit-discrete}
%&-\omega_{0}^{2}\left[\left(\phi_{n+1}-2 \phi_{n}+\phi_{n-1}\right)+\frac{c_{3}}{2!}  \left(\phi_{n+1}^{2}-2 \phi_{n}^{2}+\phi_{n-1}^{2}\right)\right.\nonumber\\
%&\quad\quad\quad\left.+\frac{c_{4}}{3!}  \left(\phi_{n+1}^{3}-2 \phi_{n}^{3}+\phi_{n-1}^{3}\right) \right]=0
\end{align}
where $c_{2}=1$, $r=C_J/C_g$  %$c_3=\tilde{\beta}\left(\phi_{ext}\right)/\tilde{\alpha}\left(\phi_{ext}\right)$, $c_4=\tilde{\gamma}\left(\phi_{ext}\right)/\tilde{\alpha}\left(\phi_{ext}\right)$, 
with $C_g$ the shunt capacitance and $C_J$ the Josephson capacitance of the SNAIL, and $\omega_0=1/\sqrt{L_0C_g}$ with $L_0=\hbar/\left[2eI_c\tilde{\alpha}\left(\phi_{{\mathrm{ext}}}\right)\right]$ and $I_c$ the critical current of the larger JJ in the SNAIL.

In the continuum approximation, the circuit equation (\ref{eq:circuit-discrete}) becomes  \cite{Ranadive2022}
\begin{align}
\frac{\partial^2\phi}{\partial t^2}\!-\!ra^2\frac{\partial^4\phi}{\partial x^2\partial t^2}\!-\!v_0^2\frac{\partial^2}{\partial x^2}\left(\phi+\frac{c_3}{2!}\phi^2+\frac{c_4}{3!}\phi^3\right)=0,\label{eq:circuit-continuous}    
\end{align}
where $v_0=a\omega_0$ with unit cell length $a$. The second (fourth order derivative) term gives rise to dispersion, while the last two terms are the third and fourth order nonlinearities, respectively (where the order terminology refers to the degree of the nonlinearity in the potential energy function). Equation (\ref{eq:circuit-continuous}) is equivalent to Eq. (7) in \cite{Ranadive2022} by replacing the phase difference coordinate $\phi$ with the node phase coordinate.

%======================
%{\section{classical background Solitons}}
\emph{Classical background solitons.}---With eventual, %analogue  
 Hawking radiation signals in mind, we wish to consider the behavior of a weak `probe' signal $\delta\phi$ (which when quantized yields Hawking
 radiation)
%(which we will  quantize in a follow-up work \cite{followup}) 
on top of a strong classical 
%(classical) 
background  $\bar{\phi}$ (which yields the analogue black hole soliton). By substituting $\phi=\bar{\phi}+\delta\phi$ into the circuit equation \eqref{eq:circuit-continuous}, we obtain the background dynamics expression 
%background*probe
%\begin{align}
%&\frac{\partial^2\bar{\phi}}{\partial t^2}-ra^2\frac{\partial^4\bar{\phi}}{\partial x^2\partial t^2}-v_0^2\frac{\partial^2}{\partial x^2}\left(\bar{\phi}+\frac{c_3}{2!}\bar{\phi}^2+\frac{c_4}{3!}{\bar{\phi}}^3\right)\nonumber\\
%&+\frac{\partial^2\delta\phi}{\partial t^2}-ra^2\frac{\partial^4\delta\phi}{\partial x^2\partial t^2}-\frac{\partial^2}{\partial x^2}v_0^2\left(1+c_3{\bar{\phi}}+\frac{1}{2}c_4{{\bar{\phi}}}^2\right)\delta\phi\nonumber\\
%&+\mathcal{O}({\delta\phi}^2)=0.\label{eq:wave_equation_order}
%\end{align}
%The first line represents the equation of 
[$\mathcal{O}(\delta \phi^0)$ term] for $\bar{\phi}$, which coincides with Eq. (\ref{eq:circuit-continuous}), %while the second  line expresses the equation of 
and the equation for the weak probe signal $\delta\phi$ affected by the background [$\mathcal{O}(\delta \phi)$ term], which we consider in the next section.
%The classical background field $\bar{\phi}$ dynamics is given  %\eqref{eq:wave_equation_order} 
%as follows:
%\begin{align}
%&\frac{\partial^2{\bar{\phi}}}{\partial t^2}\!-\!ra^2\frac{\partial^4{\bar{\phi}}}{\partial x^2\partial t^2}\!-\!v_0^2\frac{\partial^2}{\partial x^2}\left({\bar{\phi}}+\frac{c_3}{2!}{{\bar{\phi}}}^2+\frac{c_4}{3!}{{\bar{\phi}}}^3\right)=0.\label{eq:wave_equation_order-1}
%\end{align}
%where the second term represents the dispersion and the last two terms represent the nonlinearity. 
%$\phi_n(0)=A_r \sech [\sqrt{|c_3| A_r/(12r)}(n-110)]+A_l \sech[\sqrt{|c_3| A_l/(12r)}(n-190)]$ and $d\phi_n/dt|_{t=0}=A_r(1+\frac{|c_3|A_r}{6})\sqrt{|c_3|A_r/(3r)}(\sech(\sqrt{|c_3|A_r/(12r)}(x-110))^2)\tanh(\sqrt(|c_3|A_r/(12r))(x-110))-A_l(1+|c_3|A_l/6)\sqrt(|c_3|A_l/(3r))\sech [\sqrt(|c_3|A_l/(12r))(x-190)]^2\tanh [\sqrt{|c_3|A_l/(12r)}(x-190)]$

We shall now derive classical background $\bar{\phi}$ wave solutions to Eq. (\ref{eq:circuit-continuous})  that propagate along the TWPA without changing their shape, i.e., solitons.  Such a wave is obtained when the competing effects of dispersion (which broadens a wave) and nonlinearity (which sharpens a wave) are balanced. We derive a scale-invariant nonlinear evolution equation with a stationary wave solution in the vicinity of a linear approximation by using the reductive perturbation method \cite{Taniuti1968,Taniuti1969,Taniuti1974}, %When we treat perturbations in a nonlinear system, we must be careful to take into account the inherently multiple time scales. In the reductive perturbation method, we introduce different time scales through the dispersion relation as follows. The dispersion relation in our system is given as $\omega\simeq v_0k(1-ra^2k^2/2)+\mathcal{O}(a^2)$. We introduce a small parameter $\varepsilon=(ak)^2$, which represents the deviation from the leading order, linear dispersion relation. As shown in the phase of a right-propagating plane wave: 
%\begin{align}
%kx-\omega t&=\frac{1}{a}\varepsilon^\frac{1}{2}\left(x-v_0t\right)+\frac{v_0r}{2a}\varepsilon^\frac{3}{2}t,\nonumber\\
%&=\frac{1}{a}\xi+\frac{v_0r}{2a}\tau,
%\end{align}
%different time scales ($\varepsilon^{1/2}t$ and $\varepsilon^{3/2} t$) are included by different powers of $\varepsilon$ through the dispersion relation.
which employs the so-called `stretched' variables through the Gardner-Morikawa transformation: $\xi =\varepsilon^{1/2}\left(x-v_0 t\right)$ and $\tau=\varepsilon^{3/2}t$;  
%\begin{eqnarray}
%\xi &=&\varepsilon^{\frac{1}{2}}\left(x-v_0 t\right),\nonumber\\
%\tau &=&\varepsilon^{\frac{3}{2}}t, %\label{eq:GM_transformation}
%\end{eqnarray}
these correspond to slowly changing variables in relation to changes in $x$ and $t$, with $\varepsilon$ assumed small.
%Expressing as well the nonlinear terms as a series expansion in $\varepsilon$, we first 
The phase difference coordinate is expanded as follows: ${\bar{\phi}}=\varepsilon^i{{\bar{\phi}}}^{\left(1\right)}+\varepsilon^{2i}{{\bar{\phi}}}^{\left(2\right)}+\cdots$, 
%\begin{align}
%{\bar{\phi}}=\varepsilon^i{{\bar{\phi}}}^{\left(1\right)}+\varepsilon^{2i}{{\bar{\phi}}}^{\left(2\right)}+\cdots, \label{eq:phi_expansion}
%\end{align}
where $i$ is a rational number to be determined by requiring that the dispersion and nonlinear effects are balanced.
%Substituting Eqs. \eqref{eq:GM_transformation} and \eqref{eq:phi_expansion} into Eq. \eqref{eq:wave_equation_order-1}, we obtain some equations for each $\varepsilon$ order.
%\begin{align}
%&\varepsilon^{i+2}\left[2v_0\frac{\partial^2{{\bar{\phi}}}^{\left(1\right)}}{\partial\xi\partial\tau}+ra^2v_0\frac{\partial^4{{\bar{\phi}}}^{\left(1\right)}}{\partial\xi^4}\right]\nonumber\\
%&+\varepsilon^{2i+1}\left[\frac{c_3}{2}v_0^2\frac{\partial^2\left({{\bar{\phi}}}^{\left(1\right)}\right)^2}{\partial\xi^2}\right]\nonumber\\
%&+\varepsilon^{3i+1}\left[c_3v_0^2\frac{\partial^2\left({{\bar{\phi}}}^{\left(1\right)}{{\bar{\phi}}}^{\left(2\right)}\right)}{\partial\xi^2}+\frac{c_4}{6}v_0^2\frac{\partial^2\left({{\bar{\phi}}}^{\left(1\right)}\right)^3}{\partial\xi^2}\right]\nonumber\\
%&+{\mathrm{higher~order~terms  ~in}} ~\varepsilon=0.\label{eq:reductive_perturbation}
%\end{align}

Recall that the coefficients $c_3$ and $c_4$ of the nonlinear terms are controllable by varying the external magnetic flux $\phi_{{\mathrm{ext}}} $ (see Fig. \ref{fig:nonlinearity}); from now on, we consider the situation where either one or the other  of these coefficients vanishes. For the case $c_3\neq0$ and $c_4=0$, we obtain %Eq. \eqref{eq:reductive_perturbation} reduces to
\begin{align}
\frac{\partial{{\bar{\phi}}}^{(1)}}{\partial\tau}+\frac{c_3v_0}{2}{{\bar{\phi}}}^{(1)}\frac{\partial{{\bar{\phi}}}^{(1)}}{\partial\xi}+\frac{r}{2}a^2v_0\frac{\partial^3{{\bar{\phi}}}^{(1)}}{\partial\xi^3}=0   
\end{align}
for the balanced lowest order $\varepsilon^3$ dispersion and nonlinear terms, obtained by setting $i=1$. This equation is called the Korteweg-de Vries (KdV) equation \cite{KdV1895} and is known to have soliton solutions \cite{kivshar1989}. The  KdV equation can be solved by means of the inverse scattering transform method \cite{Gardner1967}. %, where the soliton solutions are obtained by solving the energy eigenvalue problem in the potential given by the shape of the initial pulse. The number of solitons corresponds to the number of bound states of the potential;
A single soliton solution is given as \cite{Gardner1967}
\begin{align}
{\bar{\phi}}_{\mathrm{KdV}}(x,t) =A\sech^2\left[\frac{1}{a}\sqrt{\frac{c_3A}{12r}}\left(x-v_st\right)\right],\label{eq:KdV-soliton}    
\end{align}
with the soliton amplitude $A$, velocity $v_s{=v_0(1+c_3 A/6)}$, and half-width $w \sim 2 a \sqrt{12 r /\left(c_{3} A\right)}$. 
%in the laboratory frame indicated in 
%[see Fig. \ref{fig:soliton} (a)]. 
The sign of the nonlinear term $c_3$ changes the polarity of the soliton; when $c_3>0$ ($c_3<0$), the soliton amplitude becomes $A>0$ ($A<0$).

For the case  $c_3{=0}$ and $c_4{\neq}0$, we obtain the %Eq. \eqref{eq:reductive_perturbation} reduces to the following, 
so-called,  `modified Korteweg-de Vries' (mKdV) equation \cite{Miura1968}:
\begin{align}
\frac{\partial{{\bar{\phi}}}^{(1)}}{\partial\tau}+\frac{c_4v_0}{4}\left({{\bar{\phi}}}^{(1)}\right)^2\frac{\partial{{\bar{\phi}}}^{(1)}}{\partial\xi}+\frac{r}{2}a^2v_0\frac{\partial^3{{\bar{\phi}}}^{(1)}}{\partial\xi^3}=0
\end{align}
for the  balanced lowest order $\varepsilon^{5/2}$ dispersion and nonlinear
terms, obtained by setting $i=1/2$. The sign of the nonlinear ($c_4$) term in the mKdV equation makes a difference to the soliton solution, in contrast to the KdV equation; when $c_4>0$, the equation is termed `mKdV$^+$' and admits a bell-shaped soliton solution given by the expression ${\bar{\phi}}_{\mathrm{mKdV^+}}(x,t)=A\sech\left[(A/a)\sqrt{\left|c_4\right|/(12r)}\left(x-v_st\right)\right]$.
%\begin{align}
%{\bar{\phi}}(x,t)=A\sech\left[\frac{A}{a}\sqrt{\frac{\left|c_4\right|}{12r}}\left(x-v_st\right)\right].\label{eq:mKdV+-soliton}    
%\end{align}
On the other hand, for $c_4<0$, the equation is termed `mKdV$^-$' and admits a shock-wave type soliton \cite{Perelman1974,Chanteur1987} given by ${\bar{\phi}}_{\mathrm{mKdV^-}}(x,t)=A\tanh\left[(A/a)\sqrt{|c_4|/(12r)}\left(x-v_st\right)\right]$.
%\begin{align}
%{\bar{\phi}}(x,t)=A\tanh\left[\frac{A}{a}\sqrt{\frac{|c_4|}{12r}}\left(x-v_st\right)\right].\label{eq:mKdV--soliton}
%\end{align}
Both soliton-types have amplitude $A$, velocity $v_s$, and half width $w \sim (2 a/A) \sqrt{12 r /|c_{4}|}$. 

%It is interesting to note 
We emphasize that both types of soliton (KdV and mKdV) are predicted to exist in the very same transmission line, which has not been considered before. This prediction is based on the reductive perturbation method, and 
%which has been widely accepted in soliton theory without verification. 
thus the SNAIL TWPA transmission line provides a unique opportunity to verify this method of analysis experimentally and understand the resulting soliton dynamics. 
%Numerical simulations of the discrete circuit equations (\ref{eq:circuit-discrete}) also validate this approximation method,  as we now show. %The above soliton solution derivations allow us to verify the experimental relevance %existence 
%of different nonlinear soliton equations obtained via the reduced perturbation method, all for the same circuit; such reduced perturbation method solutions had previously been taken for granted without experimental justification.
%, thereby contributing to the foundation of soliton physics.
%in the laboratory frame. 
%Figure \ref{fig:soliton} (b) and \ref{fig:soliton} (c) represent each soliton, respectively.
%\begin{figure}[htbp]
%\centering
%\includegraphics[width=6cm]{soliton-230120.eps}
%\caption{Schematic plots of solitons with  amplitude $A$, width $w$, and velocity $v_s$ propagating along the transmission line. (a) Soliton solution of the KdV equation. (b) Soliton solution of the mKdV$^{+}$ equation. (c) Soliton solution of the mKdV$^{-}$ equation. } \label{fig:soliton}
%\end{figure}

The above approximate, continuum field equation (m)KdV-soliton solution derivations are also verified in part by numerically solving the discrete SNAIL TWPA transmission line circuit equations \eqref{eq:circuit-discrete}. In particular, we consider the situation where the TWPA has a third order nonlinearity and set the parameter values to be $c_3=0.32$, $c_4=0$, and $r=0.1$. Setting the initial phase difference coordinate and phase velocity boundary values as $\phi_n(0)={\bar{\phi}}_\mathrm{KdV}(a (n-n_{0}),0)$ and $\dot{\phi}_n(0)=\dot{\bar{\phi}}_\mathrm{KdV}(a (n-n_{0}),0)$, 
%\begin{align}
%\phi_n(0) =A\sech^2\left[\sqrt{|c_3|A/12r}\left(n-n_0\right)\right],\label{eq:initial phi}
%\end{align}
%and 
%\begin{align}
%&\left.\frac{d\phi_n}{d\bar{t}}\right\rvert_{\bar{t}=0}=A\left(1+\frac{|c_3|A}{6}\right)\sqrt{\frac{|c_3|A}{3r}}\nonumber\\
%&\times\sech ^2\left[\sqrt{\frac{|c_3|A}{12r}}(n-n_0)\right]
%\tanh \left[\sqrt{\frac{|c_3|A}{12r}}(n-n_0)\right],\label{eq:initial dphidt}    
%\end{align}
%with the amplitude $A$, center position of the initial wave $n_0$, and $\bar{t}=\omega_0 t$,
where the dot denotes the time derivative, we confirm that the numerical solution propagates along the TWPA transmission line without changing shape %as shown in Fig. \ref{fig:propagation}, 
and is well-approximated by the analytical solution \eqref{eq:KdV-soliton} obtained by the reductive perturbation method; the numerical half-width of the soliton is about 27 unit cells when the amplitude $A = 0.02$, which is large enough for the continuum approximation solution \eqref{eq:KdV-soliton}  to be valid.
%analytical solution \eqref{eq:KdV-soliton} obtained by the reductive perturbation method propagates without changing the shape in the transmission line as shown in Fig. \ref{fig:propagation}. 
%In Fig.  \ref{fig:propagation}, the half-width of the soliton is about 27 unit cells, which is enough large for the continuum approximation and therefore the numerical simulation matches well with the analytical solution.

\begin{figure}[htbp]
\centering
\includegraphics[width=9cm]{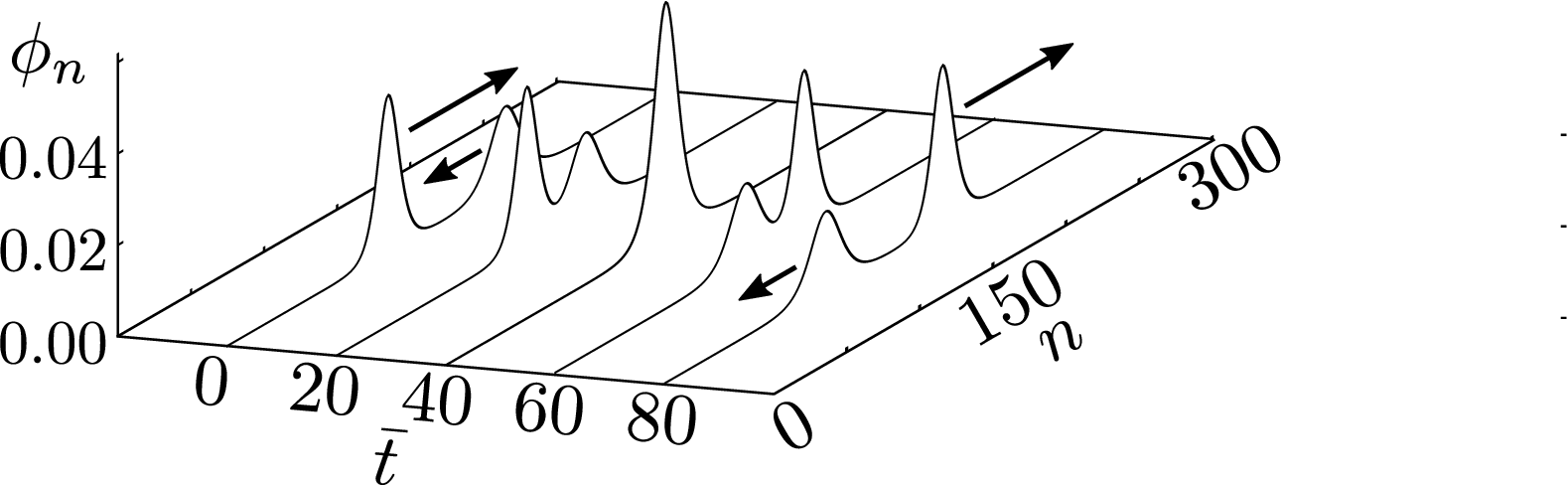}
\caption{Collision of two solitons propagating in opposite directions. The numerical solutions of the circuit equation \eqref{eq:circuit-discrete} are indicated at various time instants, where $\bar{t}=\omega_0 t$. %in each embedded diagram. 
We set the initial right (`$r$') and left (`$l$') propagating wave values to be $A^r=0.04$, $A^l=0.02$, $n^r_0=110$, and $n^l_0=190$. The circuit parameter values are given as $c_3=0.32$, $c_4=0$, and $r=0.1$. %The same circuit parameter values are used as in Fig. \ref{fig:propagation}. 
} \label{fig:collision}
\end{figure}

In order to further verify the soliton aspects of the propagating wave \cite{Zabusky1965}, we also consider two solitons initially propagating towards each other by setting the initial phase difference coordinate and phase velocity boundary values to be $\phi_n(0)=\phi^r_n(0)+\phi^l_n(0)$, $\dot{\phi}_n(0)=\dot{\phi}^r_n(0)-\dot{\phi}^l_n(0)$,  %$d\phi_n/dt|_{\bar{t}=0}=d\phi^r_n/dt|_{\bar{t}=0}-d\phi^l_n/dt|_{\bar{t}=0}$, 
where the right (left)-propagating wave and the time derivative are given by $\phi^{r(l)}_n(0)={\bar{\phi}}_\mathrm{KdV}(a (n-n^{r(l)}_{0}),0)$ and  $\dot{\phi}^{r(l)}_n(0)=\dot{\bar{\phi}}_\mathrm{KdV}(a (n-n^{r(l)}_{0}),0)$, respectively. %The center starts at SNAIL unit cell integer label location $n_0^r\, (n_0^l)$ with the amplitude $A^r\, (A^l)$. 
Figure \ref{fig:collision} shows that the later time soliton shapes are unaffected by the collision, coinciding with the earlier shapes. 
%\begin{align}
%\frac{d\phi_n}{dt}|_{t=0}=A_r(1+\frac{|c_3|A_r}{6})\sqrt{\frac{|c_3|A_r}{3r}}\sech ^2[\sqrt{\frac{|c_3|A_r}{12r}}(n-110)]\tanh [\sqrt{\frac{|c_3|A_r}{12r}}(n-110)]\\-A_l(1+\frac{|c_3|A_l}{6})\sqrt{\frac{|c_3|A_l}{3r}}\sech ^2[\sqrt{\frac{|c_3|A_l}{12r}}(n-190)]\tanh [\sqrt{\frac{|c_3|A_l}{12r}}(n-190)]   
%\end{align} 

We now briefly consider the observation of solitons in a possible experiment setup. The solitons may be detected by measuring the time variation in the voltage at the opposite end of the TWPA transmission line from the initial, injected pulse end. %When the soliton is formed %as written in either Eqs. \eqref{eq:KdV-soliton}, \eqref{eq:mKdV+-soliton}, or \eqref{eq:mKdV--soliton}, 
%we have the relation %\eqref{eq:Josephson-relation} 
%\begin{align}
%a \frac{\partial V(x,t)}{\partial x}=\frac{\hbar}{2e}\frac{\partial\phi(x,t)}{\partial t},    
%\end{align}
%the voltage is given as
%\begin{align}
%V(x,t)=\frac{1}{a} \frac{\hbar}{2 e} \int \frac{\partial \phi(x',t)}{\partial t} d x'=\frac{\hbar}{2 e}\frac{v_s}{a}\bar{\phi}(x,t),    
%\end{align}
%from the Josephson relation in the continuum limit.
%\begin{align}
%V(x,t)=\frac{\hbar}{2 e}\frac{v_s}{a}\bar{\phi}(x,t).%\frac{\hbar \omega_0}{2 e}\left(1+\frac{c_3 A}{6}\right) \tilde{\phi}(x,t)    
%\end{align}
We assume various circuit parameter values comparable in magnitude to those quoted in Ref. \cite{Esposito2022}:  $r=C_J/C_g = 0.1$, $I_c = 1.5\, \mu$A, $a=10\,\mu $m, $\alpha=0.2$, and $\phi_{{\mathrm{ext}}}=1.19 \pi$ giving $\tilde{\alpha}(\phi_{{\mathrm{ext}}})=0.37$, $c_3=0.32$, and  $c_4=0$. The voltage amplitude of the soliton with $A = 0.02$ in the phase difference coordinate %shown in Fig. \ref{fig:propagation} 
is estimated as $V=\frac{1}{a} \frac{\hbar}{2 e} \int \frac{\partial \bar{\phi}(x',t)}{\partial t} d x'=\hbar \omega_0/(2 e)\left(1+c_3 A/6\right)A=0.86\,\mu$V and the soliton temporal width is $\Delta t=w/v_s=0.21$ nsec. In our SNAIL TWPA model, solitons form spontaneously when an initial pulse is injected at one end. The soliton profile can be estimated using the initial pulse via inverse scattering theory \cite{Gardner1967}. The affect of disorder in a real array will cause some attenuation of the soliton amplitude as it propagates down the transmission line \cite{bass1988}.
The  SNAIL-TWPA is engineered such as to be impedance-matched with  $50~\Omega$ transmission lines connected at the end \cite{Esposito2022,Perelshtein2022} 
%matching should be maintained by using the resistor at the end as in 
in order to minimize reflection of the soliton  waves.  %According to inverse scattering theory, the amplitude $A$ of the soliton solution to the KdV equation is given by $A=2\lambda$, where $\lambda$ is an eigenvalue of the effective Schr\"{o}dinger equation with the initial pulse profile giving the potential energy function \cite{Gardner1967}. 
%The soliton formation process in the TWPA  from a given initial pulse will be described in our follow-up work. %\cite{followup}.

%Figure shows the formation of a soliton from an initial Gaussian voltage pulse; we assume various  circuit parameter values for existing, related SNAIL transmission line devices \cite{Frattini2017,Perelshtein2021}:  $C_g = 100\,$fF, $C_J=10\,$fF, $I_c = 1.5\, \mu$A, $a=10\,\mu $m, $\alpha=0.2$, and $\phi_{ext}=1.25 \pi$ giving $c_3=0.3$ and $c_4=0$. %Figure 
%The initial Gaussian voltage pulse is expressed as $V_1(t)=V_A\exp\left[-t^2/(2(\Delta t/2)^2)\right]$, with  assumed parameter values
%$V_A=-3.4\mu$V and $\Delta t=0.089$ nsec. 
%We assume that the pulse is equivalent to that the voltage in the transmission line has the Gaussian shape for the space with the amplitude $V_A$ and width $\Delta x=v_0 \Delta t$ at $t=0$. 

%\begin{figure}[htbp]
%\centering
%\includegraphics[width=8cm]{soliton-propagation-230120.eps}
%\caption{Propagation of a soliton numerical solution to the discrete circuit equations \eqref{eq:circuit-discrete}, plotted at various time instants for the parameter values $c_3=0.32$, $c_4=0$, and $r=0.1$. The normalized time instants $\bar{t}=\omega_0 t$ are shown above the corresponding wave peaks. The initial phase coordinate and velocity boundary values are given by Eqs. \eqref{eq:initial phi} and \eqref{eq:initial dphidt}, respectively, with the amplitude $A=0.02$ and the initial center position of the wave $n_0=150$.} \label{fig:propagation}
%\end{figure}

%In our numerical simulations, the above temporal voltage pulse is replaced with a spatial Gaussian pulse $V_1(x,0)=V_A\exp\left[-x^2/(2(\Delta x/2)^2)\right]$ that serves as the initial boundary condition with $\Delta x=v_0 \Delta t\sim 12a$, which is  wide enough for the continuum approximation to be valid. The voltage pulse changes the phase differences between junctions through Eq. \eqref{eq:Josephson-relation} and the phase difference $\phi_n$ evolves following Eq. \eqref{eq:circuit-discrete}. Figure shows the evolution of the pulse amplitude, %Figure
%becoming a soliton after around 0.3 nsec through the discharge of ripples in the direction opposite to that of the travelling pulse, and changing in width and amplitude. The numerical results agree well with the above analytical solutions. 
%The soliton can be detected by observing the voltage at the end of the transmission line. 
%Figure

\begin{figure}[htbp]
\centering
\includegraphics[width=8cm]{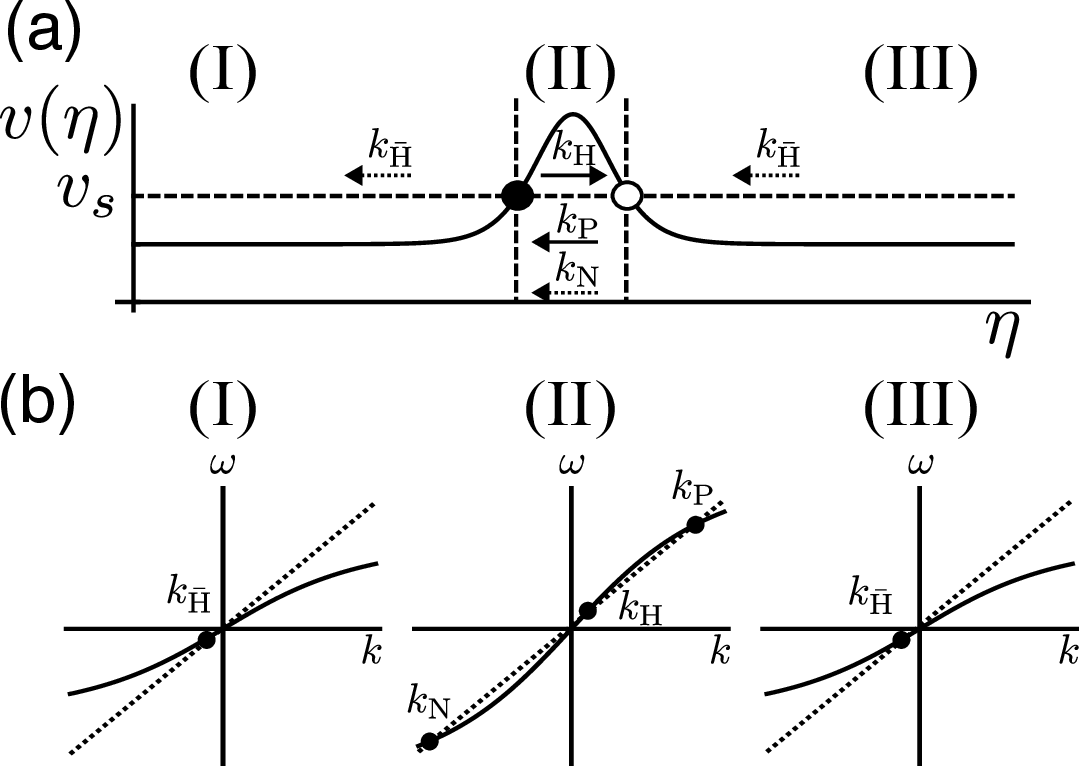}
\caption{(a) Dependence of the probe signal %microwave 
velocity in the background solition's comoving frame with spatial coordinate $\eta=x-v_s t$. The probe velocity is modulated by the soliton solution of the KdV equation. %The probe velocity profiles for the mKdV$^{+}$ and mKdV$^{-}$ look qualitatively the same. %, (b) the mKdV$^{+}$ equation, and (c) the mKdV$^{-}$ equation, resulting in analogue black-white hole horizon pairs. 
The horizontal dashed line indicates the soliton velocity. Two horizons are formed where the velocities of the probe %microwave 
and the soliton are the same, i.e., $v(\eta)=v_s$. The filled and open circles represent the horizons for the black  and white holes, respectively. Regions I and III demarcated by the vertical dashed lines correspond to the black and white hole interiors, respectively, while region II corresponds to the exterior region between the black and white hole horizons. (b) The dispersion relation of the %microwave 
probe for each region is represented by the solid line in the plot. The existing modes are indicated by the intersections with the dotted line, which shows the Doppler shift. The directions of these modes are illustrated by arrows in (a), where the solid (dotted) arrow indicates the mode with a positive (negative) frequency.}\label{fig:microwave-velocity}
\end{figure}

%======================
%\section{Analogue black-white  hole pair induced by a soliton}
\emph{Soliton as analogue black and white hole event horizons.}---We now move on to considering the behavior of a weak `probe' (Hawking radiation) signal propagating in a background soliton field, described by the 
%interacts with the obtained soliton.
$\mathcal{O}(\delta \phi)$ part of the $\phi=\bar{\phi}+\delta\phi$ substitution into Eq. \eqref{eq:circuit-continuous}: %\eqref{eq:wave_equation_order}:
\begin{align}
\left[\frac{\partial^2}{\partial t^2}-ra^2\frac{\partial^4}{\partial x^2\partial t^2}-\frac{\partial^2}{\partial x^2}v^2(x,t)\right]\delta \phi=0,\label{eq:probe}
\end{align}
where the %microwave 
probe field velocity is $v\left(x,t\right)=v_0\sqrt{1+c_3{\bar{\phi}}%(x,t)
+\frac{1}{2}c_4{{\bar{\phi}}}^2%(x,t)
}.$
%\begin{align}
%v\left(x,t\right)=v_0\sqrt{1+c_3{\bar{\phi}}(x,t)+\frac{1}{2}c_4{{\bar{\phi}}}^2(x,t)}.
%\end{align}
Note that the probe wave velocity is  modulated by the classical background soliton solution through the  space and time dependent effective inductance term $L=L_0/(1+c_3{\bar{\phi}}+\frac{1}{2}c_4{{\bar{\phi}}}^2)$. Figure \ref{fig:microwave-velocity}(a) gives  the resulting spatial dependence of the %microwave 
probe velocity in the comoving frame $\eta=x-v_s t$ traveling at the soliton velocity $v_s$. 

In order to derive the effective curved spacetime analogue of the probe wave (\ref{eq:probe}), we first transform to the comoving frame of the soliton pulse. Equation \eqref{eq:probe} then becomes $\left[-\partial_t^2+2v_s\partial_\eta\partial_t+\partial_\eta (v^2(\eta)-v_s^2)\partial_\eta\right]\delta \varphi=0,$
%\begin{align}
%\left[-\frac{\partial^2}{\partial t^2}+2v_s\frac{\partial^2}{\partial \eta\partial t}+\frac{\partial}{\partial \eta}v^2(\eta)\frac{\partial}{\partial \eta}\right]\delta \varphi=0,
%\end{align}
where the phase field coordinate $\delta \varphi$ is defined in terms of the original probe coordinate by  $\delta \phi=a (d\delta \varphi/d\eta)$. This $1+1$ dimensional wave equation is conformally invariant, which prevents us from introducing an effective, curved space-time metric. However, taking into account the two additional `inert' transverse $y,
\, z$ dimensions of the transmission line, the resulting 3+1 dimensional wave equation can be formally expressed in the following general covariant form \cite{Schutzhold2005,Blencowe2020}: $(1/\sqrt{-g})\partial_{\mu}\left(\sqrt{-g} g^{\mu \nu} \partial_{\nu} \delta\varphi\right)=0$,
%\begin{align}
%\frac{1}{\sqrt{-g}} \partial_{\mu}\left(\sqrt{-g} g^{\mu \nu} \partial_{\nu} \varphi\right)=0 ,   
%\end{align}
where $g=\det(g_{\mu\nu})$, and with effective metric given by  %Painlev\'{e}-Gullstrand form
\begin{align}
g^{\mu\nu}=\frac{1}{v(\eta)}\left(\begin{array}{cccc}
-1 & v_{s} & 0& 0\\
v_{s} & v^{2}(\eta)-v_{s}^{2} & 0 & 0\\
0 & 0 & 1& 0 \\
0 & 0 & 0& 1 \\
\end{array}\right).
\end{align}
The weak probe field can therefore be considered as propagating in an effective  curved space-time, and
an event horizon occurs at $g_{{11}}=0$, i.e., $v^2(\eta)=v_s^2$. For a background field solution corresponding to a single, right-moving classical soliton  traveling with velocity $v_s>0$, two horizons are formed as shown in Fig. \ref{fig:microwave-velocity}(a), where the {leading} and {trailing} edges of the soliton correspond to the white and black hole horizons, respectively. In the comoving frame of the soliton, the probe signal travels to the right between the horizons and to the left otherwise (also in the comoving frame). A probe signal trailing the soliton (in the lab frame) cannot reach the black hole horizon, so that the  region to the left of the black hole corresponds to its interior. Furthermore, a right moving probe signal in the neighborhood of the soliton peak (i.e., the `exterior' region in between the black and white hole horizons) cannot cross the white hole horizon, so that the region to the right of the white hole horizon corresponds to the white hole interior.

Figure \ref{fig:microwave-velocity}(b) gives the dispersion relation for a %microwave 
probe wave in the laboratory frame (curved solid lines) and the Doppler shift $\omega=\omega'+v_s k$ (dotted straight lines), where $\omega'$ is the fixed frequency in the comoving frame. The intersections indicate the existing modes in our system. In regions I and III, there is only one mode labelled $k_{\bar{\mathrm{H}}}$ which travels to the left in the comoving frame. On the other hand, in region II, the dispersion relation is modulated by the soliton and there exist three modes: one moving to the right ($k_{\mathrm{H}}$) and the other two travelling to the left in the comoving frame ($k_{\mathrm{P}}$ and $k_{\mathrm{N}}$, with `P' and `N' denoting the positive and negative frequency $\omega$ aspects). In vacuum, pair production occurs spontaneously through quantum fluctuations; photons with positive ($k_{\mathrm{H}}$) and negative ($k_{\mathrm{\bar{H}}}$) frequency $\omega$ correspond to the Hawking particles and their partners, respectively. Mode conversion ($k_\mathrm{H}+k_{\mathrm{\bar{H}}}$$\rightleftarrows$  $k_\mathrm{P}+k_\mathrm{N}$) occurs at both horizons  which behave like a cavity, resulting in lasing \cite{Corley1999,Katayama2021laser}; %This laser phenomenon is achieved through dispersion caused by the Josephson capacitor in a series of transmission lines \cite{Katayama2021laser}.
Hawking radiation is amplified as a result, which makes its observation easier.

%Finally, we discuss the observability of Hawking radiation. One way to identify an observed photon as Hawking radiation is to use its thermal properties.
%One way to determine whether the observed photons are Hawking radiation is to verify their thermal properties. 
The Hawking temperature in our system is given as $T_H=\hbar/(2\pi k_B) |\partial v(\eta)/\partial \eta|_{\eta=\eta_h}$, where $k_B$ is the Boltzmann constant and $\eta_h$ is the position of the horizon in the comoving frame \cite{Schutzhold2005}. For the parameters of the SNAIL-TWPA device investigated in Ref. \cite{Esposito2022}, $T_H\sim 0.1~{\mathrm{mK}}$, 
%\green{as previously reported}
which may be just detectable using for example a nuclear demagnetisation cryostat set-up such as that of Ref.~\cite{Cattiaux2021}. 
However, the Hawking temperature can be increased by optimizing the circuit parameters, ensuring a sufficient level of quantum fluctuations and a soliton width large enough for the continuum approximation to hold. By utilizing a different set of feasible SNAIL parameter values comparable in magnitude to those of the device investigated in Ref. \cite{Perelshtein2022}: $I_c = 1\, \mu{\mathrm{A}}$ and $r=C_J/C_g =1$, the Hawking temperature can reach several tens of milliKelvins, which is observable under ordinary dilution fridge operating temperature conditions. This enhancement occurs due to the larger capacitance ratio $r=C_J/C_g$, resulting in a larger soliton amplitude for a given width $w \sim 2 a \sqrt{12 r /\left(c_{3} A\right)}$, and increased gradient $|\partial v(\eta)/\partial \eta|_{\eta=\eta_h}$.% \blue{To distinguish Hawking radiation from other forms of thermal noise, it is important to observe the spectrum of black body radiation at the Hawking temperature. In our analogue black hole experiment, the Hawking temperature depends on the soliton amplitude, and measuring this dependence can provide evidence for the existence of Hawking radiation.}

%\begin{align}
%g^{\mu\nu}=\frac{1}{v(\eta)}\left(\begin{array}{cccc}
%-1 & v_{s} & 0 & 0\\
%v_{s} & v^{2}(\eta)-v_{s}^{2} & 0 & 0\\
%0 & 0 & 1 & 0\\
%0 & 0 & 0 & 1\\
%\end{array}\right).
%\end{align}
%\section{Summary}
\emph{Conclusion.}---%We have shown that the solitonic analogue of a black \blue{hole and} white hole pair can be realized in a TWPA transmission line comprising SNAIL elements.  The latter elements allow us to selectively obtain a third as well as fourth order nonlinearity due to the external flux tunability of the quantum phase interference. We have shown that (m)KdV soliton solutions exist in the spatial continuum field limit by using the reductive perturbation method. These analytical solutions are  validated by the numerical solution of the original discrete circuit equations. The soliton wave changes the velocity of a weak probe signal, resulting in  an effective curved space-time. Suitable SNAIL TWPAs have already been experimentally demonstrated as sources of parametrically-generated multimode, entangled microwave photons. Therefore, the experimental verification of microwave analogue black \blue{hole and} while hole pairs and Hawking radiation from the latter is a promising prospect for such SNAIL TWPA  devices. An analysis of the corresponding quantum soliton dynamics, with backreaction taken into account resulting in soliton evaporation due to Hawking radiation \blue{lasing}, will be the subject of a follow-up work. %\cite{followup}.
We have shown that (m)KdV soliton solutions exist for the circuit equations describing a TWPA transmission line comprising SNAIL elements. The soliton wave changes the velocity of a weak probe signal, describable as an effective curved space-time with analogue black hole and white hole event horizons. Suitable devices have already been experimentally demonstrated as sources of parametrically-generated multimode, entangled microwave photons. Therefore, the experimental verification of microwave analogue black hole and white hole pairs and Hawking radiation is a promising prospect. An analysis of the corresponding quantum soliton dynamics, with backreaction taken into account resulting in soliton evaporation due to Hawking radiation lasing, will be the subject of a follow-up work. 

\begin{acknowledgments}
%\section{Acknowledgements}
We thank Maxime Jacquet, Frieder Koenig, Anja Metelmann, Grigory Volovik, Pertti Hakonen, Alexander Zyuzin, Kirill Petrovnin, Satoshi Ishizaka, and Seiji Higashitani for very helpful discussions. We also thank Terry Kovacs for computational assistance. MPB was supported by the NSF under Grant No. PHY-2011382. HK was supprted by Marubun Exchange Research Grant and JSPS KAKENHI 22K20357. NH and TF were supported by JSPS KAKENHI 22K03452. 
\end{acknowledgments}

%\end{CJK}
\bibliography{hoge-resub}
%\bibliographystyle{junsrt} 
%\bibliographystyle{apsrev4-2}
%=====

\end{document}